\documentstyle[12pt]{article}
\begin{document}
\vskip 2 cm
\begin{center}
\Large{\bf WHAT IS THE STANDARD MODEL HIGGS ? }
\end{center}
\vskip 3 cm
\begin{center}
{\bf AFSAR ABBAS} \\
Institute of Physics, Bhubaneswar-751005, India \\
(e-mail : afsar@iopb.res.in)
\end{center}
\vskip 20 mm  
\begin{centerline}
{\bf Abstract }
\end{centerline}
\vskip 3 mm

It is shown that in the Standard Model, the property of charge
quantization holds for a Higgs with arbitrary isospin and hypercharge.
These defining quantum numbers of the Higgs remain unconstrained while the
whole basic and fundamental structure of the Standard Model remains
intact. Hence it is shown that the Higgs cannot be a physical particle.
Higgs is the underlying 
`vacuum' over which the whole edifice of the Standard Model stands.

\newpage

The Standard Model (SM) of particle physics is a well-tested and
established model. Higgs boson is the only missing link. Much effort, both
theoretical and experimental (for a review see [ 1 ]) worldwide is being
put to understand and detect it. It is a common belief that future
colliders like LHC shall detect it in the near future. But do we
understand Standard Model well enough to speak with confidence of the
impending discovery of Higgs boson ? 

 The answer is No ! Until 1989/90 it was commonly believed (unfortunately
many still believe it) that the SM does not have electric charge
quantization built into it and that one has to go to the Grand Unified
Theories (GUTs) to obtain charge quantization [ 2 ]. In fact one of the
main motivations of invoking GUTs was precisely this. But it was
shown that electric charge quantization is built into the SM [ 3,4,5,6 ].
However, some authors were making redundant assumptions of arbitrarily
fixing up the Higgs hypercharge in these derivations [ 3,4 ]. This
weakened their charge quantization arguments and was corrected by the
author [ 5,6 ]. Pitfalls in such weakening assumptions were also pointed
out by the author [ 6 ]. Until very recently, it was believed that the SM
allows for the existence of the millicharged particles. This is not true
[ 6 ] and the point was clarified in a recent letter by the author [ 7 ].

 So has the last hurdle in understanding the SM been crossed ? No, because
we have not yet found the Higgs. In this paper, the author would show that
this is due to deeper reasons. We still have not understood the basic
underlying structure 
of the SM and hence it is very relevant to ask, what is the standard 
model Higgs ?

 The SM assumes a repetitive structure for each generation of quarks and
leptons. The anomalies cancel generation by generation. Why this is so, is
not answered by the SM. We accept this as a fact (any attempt to do so,
would take one beyond the SM) and try to understand the deep and basic
underlying structure of the SM.

Let us start by looking at the first generation of quarks and leptons
(u, d, e,\,$\nu$ )  and assign them to
$SU(N_{C}) \otimes SU(2)_L \otimes U(1)_Y$ (where $ N_{C} 
= 3 $ ) representation as follows.

\begin{displaymath}
q_L = \pmatrix{u \cr d}_L, (N_{C},2,Y_q)
 \end{displaymath}
\begin{displaymath} u_R; (\bar{ N_{C} },1,Y_u) \end{displaymath}
\begin{displaymath} d_R; (\bar{N_{C}},1,Y_d) \end{displaymath}
\begin{displaymath} l_L =\pmatrix{\nu \cr e}; (1,2,Y_l)
\end{displaymath}
\begin{equation}
 e_R = (1,1,Y_e)
\end{equation}

To keep things as general as possible this brings in five unknown
hypercharges.Let us now define the electric charge in the most general
way in
terms of the diagonal generators of $SU(2)_L \otimes U(1)_Y$ as
\begin{equation} Q'= a'T_3 + b'Y \end{equation}
\newline We can always scale the electric charge once as $Q={Q'\over
a'}$ and hence ($b={b'\over a'}$)
\begin{equation} Q = T_3 + bY \end{equation}

In the SM $ SU(N_{C}) $ $\otimes$ $ SU(2)_{L}$ $\otimes$
$U(1)_{Y}$ is spontaneously broken through the Higgs mechanism to the
group $ SU(3)_{c} $ $\otimes$ $U(1)_{em}$ . In SM the Higgs is
assumed to be a doublet. However we do not use this restriction either and
assume the Higgs $ \phi $ to have any isospin T and arbitrary hypercharge
$ Y_{\phi} $.
 The isospin $ T_{3}^{\phi} $ component of the
Higgs develops a nonzero vacuum expectation value $<\phi>_o$. Since we
want
the $U(1)_{em}$ generator Q to be unbroken we require $Q<\phi>_o=0$. This
right away fixes b in (3) and we get
\begin{equation} 
Q = T_3 - ( \frac{ T_{3}^{\phi} }{ Y_{\phi} } ) Y
\end{equation}

For the SM to be renormalizable we require that the triangular anomaly
be canceled. This leads to three constraints.

\begin{eqnarray}
Tr Y [ SU( N_{C} ) ]^{2} = 0 \\
Tr Y [ SU(2)_{L} ]^{2} = 0 \\
Tr [ Y^{3} ] = 0 
\end{eqnarray}

The expression (6) yields 

\begin{equation}
Y_{q} = - \frac{ Y_{l} }{ N_{ C } }
\end{equation}

Next we use the fact that after the spontaneous breaking of 
$ SU( N_{ C } ) \otimes SU( 2 )_{L} \otimes U( 1 )_{ Y } $ to 
$ SU( N_{C} ) \otimes U(1)_{ em } $, the L- and R-handed charges couple
identically with photon. Using 
$ Q ( u_{L} ) = Q ( u_{R} ); Q ( d_{L} ) = Q ( d_{R} ); 
Q ( e_{L} ) = Q ( e_{R} ) $ yields
respectively

\begin{eqnarray}
Y_{u} = Y_{q} - \frac{ Y_{ \phi } }{ 2 T_{3}^{ \phi } } \\
Y_{d} = Y_{ q } + \frac{ Y_{\phi} }{ 2 T_{3}^{\phi} } \\
Y_{e} = Y_{l} + \frac{ Y_{\phi} }{ 2 T_{3}^{\phi} } 
\end{eqnarray}

 When these are used in conjunction with the condition (7) one finds

\begin{equation}
Y_{l} = \frac{ Y_{\phi} }{ 2 T_{3}^{\phi} }
\end{equation}

Hence

\begin{equation}
Y_{q} = - \frac{ Y_{\phi} }{ 2 T_{3}^{\phi} N_{C} }
\end{equation}

When these hypercharges are put in eq. (4) for the electric charge one
obtains :

\begin{displaymath} 
\newline Q(u) = {1\over 2}(1+{1\over N_c})
\end{displaymath}
\begin{equation}  
\newline Q(d) = {1\over 2}(-1+{1\over N_c})
\end{equation}

\begin{eqnarray}
Q(e) = - 1 \\
Q( \nu ) = 0
\end{eqnarray}

For $ N_{C} = 3 $ these are the correct charges in the SM. Note that this
charge quantization in the SM holds for Higgs for arbitrary T and
arbitrary hypercharge. Hence as far as charge quantization is concerned,
the values of T and $ Y_{ \phi } $ remain unconstrained. This point for
the special case of the Higgs doublet 
was already noted by the author earlier [ 5,6 ].

 Let us continue with the rest of the structure of the SM and see how our
general Higgs with unconstrained and unspecified isospin T and
hypercharge $
Y_{ \phi } $ fits into it. We can write the covariant derivative of the SM
as

\begin{equation}
D_{\mu} = \partial_{\mu} + i g_{1} \frac{ T_{3}^{\phi} }{ Y_{\phi} }
Y B_{\mu } - i g_{2} \vec{T} . \vec{ W_{\mu} }
\end{equation}


As the kinetic energy term for the gauge fields is

\begin{equation}
- \frac{1}{4} W_{a}^{ \mu \nu } W_{ a \mu \nu } 
- \frac{1}{4} B^{ \mu \nu } B_{ \mu \nu }
\end{equation}

Only orthogonal combination of $ W_{ \mu }^{0} $ and $ B_{ \mu }^{0} $
would have independent kinetic energy terms. Hence the photon field $ A_{
\mu } $ and the orthogonal $ Z_{ \mu } $ are written as 

\begin{equation}
A_{ \mu } = \frac{ g_{ 2 }B_{ \mu } + g_{1} 
                 ( \frac{ 2 T_{3}^{ \phi } }{ Y_{ \phi } } Y_{l} )
                 W_{ \mu}^{0} }
             { \sqrt{ g_{ 2 }^{2} + 
                ( g_{1} \frac{ 2 T_{3}^{\phi} }{ Y_{\phi} }  Y_{l} )^2
               } }    
\end{equation}

\begin{equation}
Z_{ \mu } = \frac{ -g_{1} 
         ( \frac{ 2 T_{3}^{ \phi } }{ Y_{\phi} } Y_{L} ) B_{ \mu } 
             + g_{2} W_{ \mu }^{ 0 } }
             { \sqrt{ g_{2}^{2} + 
                (g_{1} \frac{ 2 T_{3}^{ \phi } }{ Y_{\phi} } Y_{l} )^{2} }
}
\end{equation}


with $ Y_{l} $ from eq. (12) we get

\begin{equation}
A_{\mu} = \frac{ g_{2} B_{ \mu } + g_{1} W_{ \mu }^{ 0 } 
             }{ \sqrt{ g_{1}^{2} + g_{2}^{2} } }
\end{equation}

and 

\begin{equation}
 Z_{ \mu } = \frac{ -g_{1} B_{ \mu } + g_{2} W_{ \mu }^{0} }
             { \sqrt{ g_{1}^{2} + g_{2}^{2} } }
\end{equation}

With $ D_{ \mu } $ given by eq.(17) we can write the lepton part of the SM
Lagrangian as 

\begin{eqnarray}
{ \cal{L} }(lepton) = \bar{ q_{L} } i \gamma^{ \mu } 
      ( i g_{1} \frac{ T_{3}^{\phi} }{ Y_{ \phi } } Y_{ l } B_{ \mu } )
       q_{L} 
       + \bar{ e_{R} } i \gamma^{ \mu } 
      ( i g_{1} \frac{T_{3}^{ \phi }}{ Y_{ \phi } } Y_{e} B_{ \mu } )
        e_{R} \nonumber \\
        - \bar{ q_{L} } i \gamma^{ \mu } 
          \left[ i g_{2} \vec{T} . \vec{ W_{\mu} } \right] q_{L}   
\end{eqnarray}

 The electron term in the Lagrangian with $ B_{ \mu }, W_{ \mu }^{0} $
replaced from eqs. (21) and (22) becomes

\begin{eqnarray}
A_{ \mu } \left[ \bar{ e_{L} } \gamma^{ \mu } e_{ L }
              ( \frac{ G_{1} g_{2} Y_{l} }{ X } 
           + \bar{ e_{R} } \gamma^{ \mu } e_{ R }
              ( \frac{ G_{1} g_{2} Y_{e} }{ 2 X } ) \right] 
\nonumber \\
 + Z_{ \mu } \left[ \bar{ e_{L} } \gamma^{ \mu } e_{L}
        ( \frac{ G_{1}^{2} Y_{l}^{2} - g_{2}^{2} }{ 2 X } )
      + \bar{e_{R} } \gamma^{ \mu } e_{ R }
        ( \frac{ G_{1}^{2} Y_{e} Y_{l} }{ 2X } ) \right]
\end{eqnarray}

 where 
$ G_{1} = - \frac{ 2 g_{1} T_{3}^{ \phi } }{ Y_{\phi} } $
and
$ X = \sqrt{ g_{2}^{2} + G_{1}^{2} Y_{l}^{2} } $

As the electromagnetic interaction of particles with charge Q is 

\begin{equation}
{ \cal{L} }_{em} = Q A_{\mu} \left[ \bar{ e_{L} } \gamma^{ \mu } e_{L}
     + \bar{ e_{R} } \gamma^{ \mu } e_{ R } \right]
\end{equation}

This form above is true provided
$ Y_{e} = 2 Y_{l} $. This is exactly what we had found above in the
context of the electric charge quantization.

As we require $ Q = - e $ for electron we find

\begin{equation}
e = \frac{ g_{1} g_{2} }{ \sqrt{ g_{1}^{2} + g_{2}^{2} } }
\end{equation}

with 

\begin{equation}
g_{2} = \frac{e}{ sin \theta_{W} },
g_{1} = \frac{e}{ cos \theta_{W} }
\end{equation}

We also find that electron coupling to $ Z_{ \mu } $ and neutron coupling
to $ Z_{ \mu } $ are as in the SM.

The point to be emphasized is that the whole structure of the SM stands
and is independent of Higgs isospin and hypercharge, which all
throughout remain
unconstrained and undetermined. One should not fix any arbitrary values
for them as nothing in the theory demands it.

We also find that 

\begin{equation}
\rho = \frac{ m_{W}^{2} }{ m_{Z}^{2} Cos^{2} (\theta_{W}) }
= \frac{ T(T + 1 ) - { T_{3}^{\phi} }^{2} }
      { 2 {T_{3}^{ \phi } }^{2} }
\end{equation}

$ \rho = 1 $ when 

\begin{equation}
T(T + 1 ) - 3 T_{3}^{ \phi } = 0 
\end{equation}

$ T = 1/2. T = \pm 1/2 $ does satisfy (29). Next solution is $ T = 3, 
T_{3}
\pm 2 $ . Actually, as pointed out by Tsao [ 8 ] the complete set of
solutions of (29) is infinite. One can generate all solutions from the
lowest doublet solution by rewriting (29) as

\begin{equation}
X^{2} - 3 Y^{2} = 1
\end{equation}

where $ T = \frac{ X - 1 }{2}, T_{3} = \pm \frac{Y}{2} $. This is a
special case of the Pell equation in number theory. All the solutions ( $
X_{n}, Y_{n} ), \, n=1,2,... $ can be obtained from integer solutions of 

\begin{equation}
X_{n} + \sqrt{3} Y_{n} = ( 2 + \sqrt{3} )^{n}
\end{equation}

The solutions of this are infinite in number. Hence for $ \rho = 1, T, 
T_{3} $ are infinite in number. Again, nothing in theory demands that one
fix this to a particular value.

 The point is that the full structure of the SM stands intact without
constraining the quantum numbers isospin and/or the hypercharge of the
Higgs to any specific value. All the
particles that have been isolated in the laboratory or have been studied
by any other means, besides having a specific mass, have definite
quantum numbers which identify them.
In the case of Higgs here, no one knows
of its mass and more importantly its quantum numbers like isospin and
hypercharge, as shown above, are not specified. The
hypercharge of all the other particles are specified as being proportional
to the Higgs hypercharge  which itself remains unconstrained. 
That is, all the hypercharges of particles are rooted on to the Higgs
hypercharge which itself remains free and unspecified. Hence Higgs is
very different from all known particles.
Because of the above reasons Higgs cannot be a physical particle which may
be isolated and studied. It must be just the `vacuum' which sets up the
structure of the whole thing. 

Higgs is a manifestation of the vacuum structure of the SM. Higgs shall
never get pinned down as an isolated physical particle, but makes
its presence felt through charge quantization and giving the SM its
complete structure and consistency. Hence it is predicted
that Higgs shall not be discovered as a particle.

 No basic principle demands that the mass of the matter particle be given
by Yukawa interaction, but since as we have no idea of where these masses
come from, one just demands that they arise from such a coupling. If this
be so then the Higgs isospin is necessarily T = 1/2. This just tells you
that the `vacuum' has this particular structure. But as $ Y_{ \phi } $ is
not constrained in any way, the Higgs cannot be a particle but just
`vacuum' which behaves in this fundamental and basic manner.

 In conventional SM there is just one Higgs doublet and which gives 
$ e- \mu - \tau $ universality for the 3 generations. In our present
calculations there is no reason that Higgs structure for each generation may be
the same. There could be three different Higgs `vacuum' structures and yet
give $ e - \mu - \tau $ universality. Instead of being a static Higgs the
Higgs in our view is a more dynamic `vacuum'.
In our model, as lack of Yukawa coupling implies unconstrained isospin of
the Higgs, hence custodial symmetry question becomes doubtful.
So should monopoles as well. The fact
that Higgs is not a particle
should also help in solving the cosmological constant problem.

 In summary, we have shown that the basic and fundamental structure of the
standard model stands intact without specifying and constraining the
quantum numbers of the Higgs. As such Higgs is very different from any
known physical particle. Hence Higgs cannot be a `particle' but represents
the omnipresent vacuum which provides the `root' to support the Standard
Model.

\newpage

\begin{center}
{\bf\large REFERENCES }
\end{center}

\vskip 2 cm
1. G.F.Gunion, H.E.Haber, G.Kane and S. Dawson, 
"The Higgs Hunter's Guide", Addison-Wesley Pub. Co., Redwood, California,
USA (1990).

2. P.D.B.Collins, A.D.Martin and E.J.Squires, 
"Particle Physics and Cosmology", John Wiley \& Sons, New York, USA (1989)

3. K.S.Babu and R.N.Mohapatra, 
{\it Phys. Rev. Letts.} {\bf 63} (1989) 938

4. X.-G.He, G.C.Joshi, H.Lew, B.H.J.McKellar and R.R.Volkas,
{\it Phys. Rev. D} {\bf 40 } (1989) 3140

5. A. Abbas, 
{\it Phys. Lett. } {\bf 238 B} (1990) 344

6. A. Abbas,
{\it J. Phys. G. } {\bf 16 } (1990) L163

7. A. Abbas,
{\it Physics Today }, July 1999, p.81-82

8. H.-S.Tsao, 
Proc. 1980 Guangzhou Conf. on Theo. Part. Phys.; Ed. H.Ning and T.
Hung-Yuan,
(Science Press, Beijing, 1980), p.1240

\end{document}